\documentclass[twocolumn]{aastex701}

\usepackage{graphicx}  
\usepackage{amsmath}    
\usepackage{amssymb}    
\usepackage{makecell}	
\usepackage{comment}

\def\cha        {{Chandra}\/}
\def\xmm        {{XMM-Newton}\/}

\begin{document}

\title{Spiraling into the `snail' cluster MCXC~J0157.4-0550}

\author[0000-0002-5996-0167]{Chao Geng}
\affiliation{Department of Astronomy, University of Science and Technology of China, Hefei 230026, People’s Republic of China}
\affiliation{School of Astronomy and Space Science, University of Science and Technology of China, Hefei 230026, People’s Republic of China}
\email{gengchao@ustc.edu.cn}

\author[0000-0003-0628-5118]{Chong Ge}
\affiliation{Department of Astronomy, Xiamen University, Xiamen, Fujian 361005, People’s Republic of China}
\email[show]{chongge@xmu.edu.cn}

\author[0000-0001-5880-0703]{Ming Sun}
\affiliation{Department of Physics and Astronomy, University of Alabama in Huntsville, Huntsville, AL 35899, USA}
\email{ms0071@uah.edu}

\author[0000-0001-7500-0660]{Li Ji}
\affiliation{Key Laboratory of Dark Matter \& Space Science, Purple Mountain Observatory, CAS}
\email{fakeemail1@google.com}

\author[0000-0003-4286-5187]{Guilin Liu}
\affiliation{Department of Astronomy, University of Science and Technology of China, Hefei 230026, People’s Republic of China}
\affiliation{School of Astronomy and Space Science, University of Science and Technology of China, Hefei 230026, People’s Republic of China}
\email{fakeemail1@google.com}

\author[0000-0003-3667-1060]{Zhicheng He}
\affiliation{Department of Astronomy, University of Science and Technology of China, Hefei 230026, People’s Republic of China}
\affiliation{School of Astronomy and Space Science, University of Science and Technology of China, Hefei 230026, People’s Republic of China}
\email{fakeemail1@google.com}

\author[0000-0001-9603-7673]{Haiguang Xu}
\affiliation{School of Physics and Astronomy, Shanghai Jiao Tong University, 800 Dongchuan Road, Minhang, Shanghai 200240, People's Republic of China}
\email{fakeemail1@google.com}

\author[0000-0003-3175-2347]{John ZuHone}
\affiliation{Center for Astrophysics | Harvard and Smithsonian, 60 Garden Street, Cambridge, MA 02138, USA}
\email{}

\author[0000-0003-0297-4493]{Paul Nulsen}
\affiliation{ICRAR, University of Western Australia, 35 Stirling Hwy, Crawley, WA 6009, Australia}
\affiliation{Center for Astrophysics | Harvard and Smithsonian, 60 Garden Street, Cambridge, MA 02138, USA}
\email{}

\author[0000-0003-0167-0981]{Craig Sarazin}
\affiliation{Department of Astronomy, University of Virginia, Charlottesville, VA, USA}
\email{}

\author[0000-0003-2898-0728]{Nobuhiro Okabe}
\affiliation{Physics Program, Graduate School of Advanced Science and Engineering, Higashi Hiroshima Campus, Hiroshima University, Hiroshima 739-8526, Japan}
\email{}

\begin{abstract}
We present a deep \cha\ (130 ks) study of a striking, tidally induced cold front in a subcluster merging with the galaxy cluster MCXC J0157.4-0550. The subcluster, nicknamed the `snail' (J0157), displays an extensive spiral structure that extends to $\sim 260$ kpc --- one of the largest spiral cold fronts identified to date. Surface brightness and spectral analysis across four sectors reveal a consistent density jump of a factor of $\sim 3$ across the front, together with significant temperature discontinuities, and thermodynamic mapping further delineates the coherent, low-entropy structure of the stripped gas. Unlike classical sloshing cold fronts, the spiral in J0157 arises not from oscillation of a displaced cool core, but from subcluster gas that has been ram pressure stripped and tidally wound into a spiral by orbital rotation. Applying Gaussian Gradient Magnitude filtering and azimuthal surface-brightness analysis, we identify quasi-periodic corrugations along the front consistent with candidate Kelvin-Helmholtz (KH) instabilities, with a characteristic scale of $\sim 90$ kpc. The presence of these KH rolls constrains the effective ICM viscosity to $\lesssim 0.59$ times the Spitzer value and the magnetic field strength at the interface to $\lesssim 5.9\ \mu$G. These results establish J0157 as a useful laboratory for probing macro-scale merger-induced gas dynamics and micro-scale plasma transport processes in the intracluster medium.
\end{abstract}


\section{Introduction}

	\begin{figure*}
		\centering 
		\includegraphics[width=2\columnwidth]{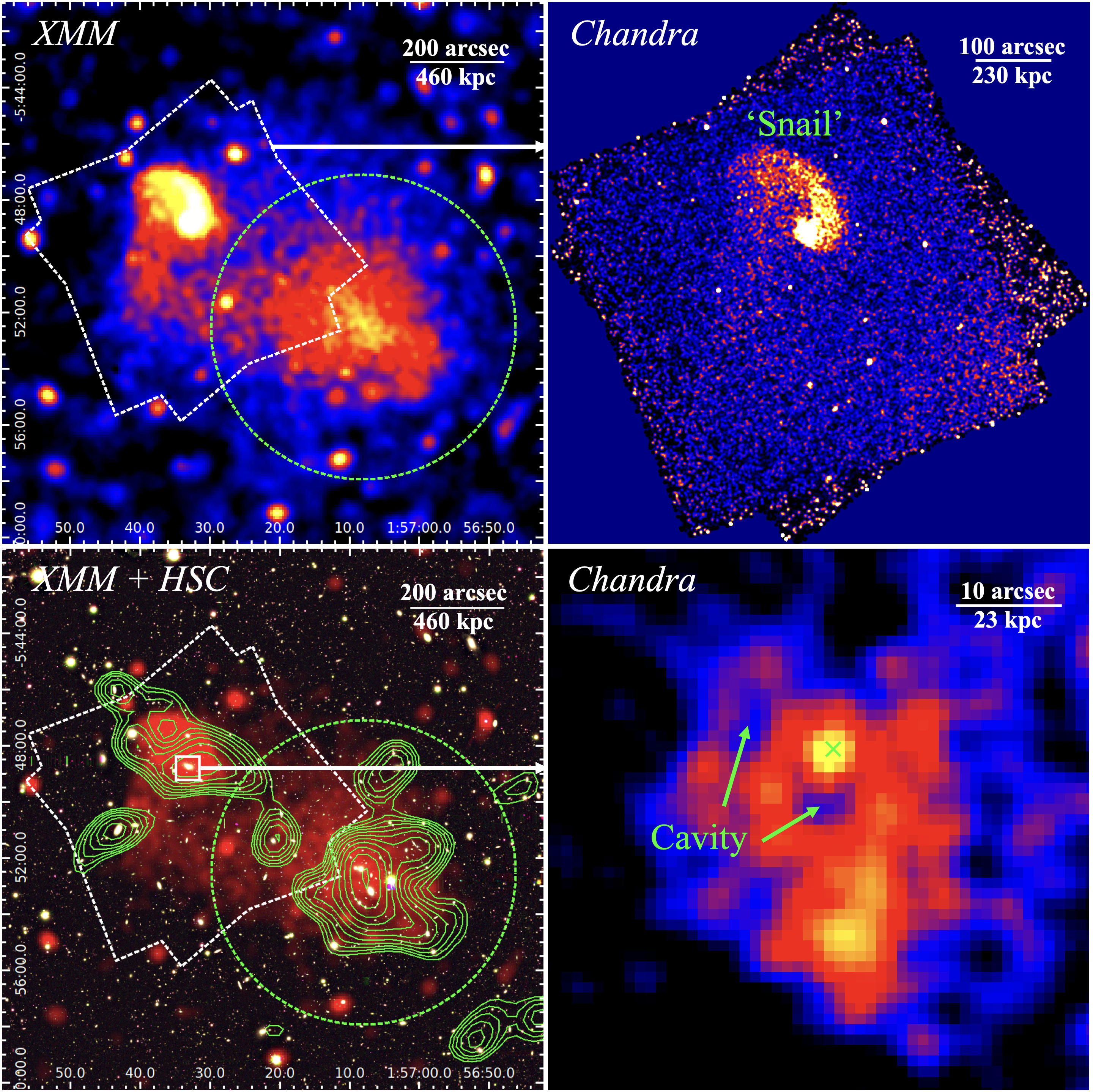}
		\caption{
        Top left: The \xmm\ 0.5-2 keV image of the merging galaxy cluster MCXC J0157.4-0550. The dashed green circle shows $R_{500}$ (750 kpc or 5.4 arcmin) of the main cluster. The `snail' shaped subcluster is to the northeast. The white dashed line shows the field of view (FOV) of three \cha\ observations (top right panel).
        Top right: The merged \cha\ 0.7-2 keV image of the snail subcluster J0157, with background subtracted and exposure corrected. The \cha\ image pixel size is 0.492 arcsec, smoothed with a uniform Gaussian kernel of 2.95 arcsec. 
        Bottom left: HSC optical image ($riz$ band) overlaid with \xmm\ image in red from above. The green contours are from the cluster galaxy distribution map. The white box shows the region of the bottom right zoomed-in image.
        Bottom right: The zoomed-in \cha\ image of the BCG (SDSS J015733.03-054843.5) of the subcluster J0157. The green cross marks the center of the BCG, with green arrows pointing to the cavity structures.
		}
		\label{fig:fig1}
	\end{figure*}

Hierarchical structure formation theory predicts that large halos form through successive mergers of subhalos. 
Extensive \cha\ observations of merging galaxy clusters have revealed characteristic cold front (CF) features, marked by sharp discontinuities in surface brightness and temperature \citep[e.g.][]{Markevitch2007}. 
These CFs typically maintain approximate pressure equilibrium across their interfaces, despite the significant jumps in other thermodynamic properties \citep[e.g.][]{Zuhone2016}.
Besides CFs in merging clusters, another type,  known as sloshing CFs, can be found in relaxed clusters, when their cool cores (CCs) are displaced and oscillate (`slosh') around the cluster center.
The latter phenomenon is triggered when a passing subcluster gravitationally perturbs the CC, displacing it from the cluster's potential minimum \citep[e.g.][]{Markevitch2001}. 
Subsequently, the offset CC falls back toward the potential minimum and thus oscillates (or sloshes) within the cluster center.
This sloshing motion can produce a spiral pattern, particularly when the passing subcluster has a non-zero impact parameter, leading to angular momentum transfer \citep[e.g.][]{2006ApJ...650..102A}.
However, these spiral patterns are usually subtle and become significant only after removing the average CC image (e.g. \citealt{2010A&A...511A..15L}).
Sloshing cores may also generate turbulence, potentially accelerating and dispersing relativistic electrons to form radio minihalos (e.g. \citealt{2013ApJ...762...78Z,Brienza2022,Giacintucci2024}). 

As summarized by \cite{Markevitch2007}, CFs serve as excellent probes for studying microtransport processes (e.g. viscosity, magnetic draping, heat conduction, and plasma mixing) in the magnetized intracluster medium (ICM). 
Due to shear flows between gas of different velocities, CFs are expected to be distorted by Kelvin-Helmholtz (KH) instabilities, which can be suppressed by magnetic fields or damped by viscosity. 
Both smooth and distorted CFs have been observed (e.g. \citealt{2000MNRAS.317L..57E,Vikhlinin2001,2001ApJ...551..160V}; \citealt{Ichinohe2017}), indicating that the KH instabilities are suppressed in some clusters but not others. 

From existing studies (e.g. \citealt{2010A&A...511A..15L}), clusters such as A1644 \citep{2010ApJ...710.1776J}, NGC~7618/UGC~12491 \citep{Roediger2012}, the Ophiuchus cluster \citep{2016MNRAS.460.2752W}, and the Perseus cluster \citep{Walker2017} show significant spiral sloshing features even without image subtraction.
In these systems, sharp sloshing edges are detected at distances up to $\sim 100$ kpc from their CC centers.
However, the SDSS J015733.03-054843.5 subcluster (hereafter J0157, Fig.~\ref{fig:fig1}) exhibits a remarkable spiral pattern, with its spiral front extending from the subcluster center up to $\sim$ 260 kpc, resembling a giant `snail' in the sky, as it falls into the main cluster MCXC J0157.4-0550 (e.g. \citealt{Miyaoka2018,Yang2026}). 
This spiral front appears particularly smooth and well-defined in the \xmm\ data.
The spiral pattern then turns sharply towards the south, with a tail detected to at least 700 kpc. 

We note that both the spiral pattern observed in J0157 and those seen in classical sloshing CFs originate from the transfer of orbital angular momentum associated with a non-zero impact parameter. However, the physical nature of the two systems is different. In classical sloshing CFs, the CC of the main cluster is displaced from the potential minimum and subsequently oscillates or sloshes within the main cluster's own gravitational potential well. The spiral morphology arises from the radial dependence of the sloshing frequency combined with the angular momentum acquired during the encounter. In J0157, by contrast, the spiral is formed by gas that has been stripped from the subcluster, and dragged by tidally induced rotation \citep{Yang2026}. This gas does not oscillate within the main cluster's potential well. Therefore, although both phenomena share a common origin in orbital angular momentum and exhibit a similar spiral appearance, J0157 differs fundamentally from the classical sloshing CFs; it is more accurately described as a merger induced or tidally driven CF.

In this paper, we present results from deep \cha\ observations of this spiral CF.
We assume a cosmology with $H_0 = 70\ {\rm km\ s}^{-1}\ {\rm Mpc}^{-1}$, $\Omega_{\rm m} = 0.3$, and $\Omega_{\Lambda} = 0.7$. 
At J0157's redshift of $z = 0.1287$, 1 arcsec = 2.3 kpc.

\section{data analysis}
\label{sec: data analysis}

The \xmm\ X-ray data in Fig.~\ref{fig:fig1} are reduced with the Extended Source Analysis Software (ESAS), integrated into the \xmm\ Science Analysis System (SAS; version 17.0.0), following \cite{2019MNRAS.484.1946G}. The observation (ID: 0781200101, PI: Nobuhiro Okabe) has clean exposure times of 27.6 ks (MOS1), 27.1 ks (MOS2), and 16.2 ks (pn).

The \cha\ X-ray observations are performed with the \cha\ Advanced CCD Imaging Spectrometer (ACIS). Three observations (ObsIDs: 22645, 22863, and 22864; PI: Ming Sun) have exposure times of 48.4, 49.4, and 32.6 ks, respectively.
We use the \cha\ Interactive Analysis of Observation (CIAO, version 4.11) and calibration database (CALDB, version 4.7.7) to reprocess the \cha\ data following procedures described in \cite{2009ApJ...693.1142S} and \cite{2018MNRAS.481.4111G}. All observations are taken in the very faint (VFAINT) mode, and we reprocess the new level 2 event files using the {\sc chandra\_repro} script with the VFAINT mode correction.
We examine the light curve in the band 0.5 $\sim$ 9.5 keV from source-free regions to remove the intervals affected by flares, using the {\sc deflare} script. Such intervals are defined as those deviating by more than 3$\sigma$ from the mean rate.
These \cha\ observations are largely unaffected by flares, with a total clean exposure of 130 ks.

We use the {\sc wavdetect} script to detect point sources with the point spread function (PSF) map created by the {\sc mkpsfmap} script, and the result is then checked through visual inspection.
For each observation, we construct the stowed background in VFAINT mode and reproject it to match that observation. The count rate of the background is rescaled based on the 9.5-12 keV counts, where the \cha\ effective area is negligible and the flux is dominated by the particle background.
We make a merged image from the three observations using {\sc merge\_obs} on the level-2 event files. After subtracting the normalized background image and dividing by the merged exposure map, we produce the exposure-corrected and background-subtracted image (Fig.~\ref{fig:fig1} top right panel).

The spectra are extracted with {\sc specextract} following \cite{Ge2021b}. For the spectral fitting of the main cluster, we first subtract the instrumental background and then subtract the sky background. We use the stowed background to subtract the instrumental background. 
For the diffuse X-ray sky contribution, we use the HEASARC X-ray Background Tool to extract an off-cluster ROSAT all-sky survey (RASS) spectrum in a region around the cluster. Then we jointly fit the background-subtracted on-source and RASS off-source spectra with the model of {\tt CONSTANT*(APEC + (APEC + POWERLAW)*TBABS + APEC*TBABS)}. The {\tt CONSTANT} parameter scales the spectral extraction area between the on-source and off-source spectra. The off-source spectrum is fitted with three components: an unabsorbed thermal model {\tt APEC} (for the Local Bubble contribution), an absorbed {\tt APEC} (the Galactic halo gas), and an absorbed {\tt POWERLAW} (the extragalactic background). Beyond these three components, an additional absorbed {\tt APEC} (representing the cluster emission) is added to fit the on-source spectrum for the host cluster. 
For the spectral fitting of the subcluster, we use the local background to remove the contamination from the main cluster diffuse gas. As the sky and main cluster backgrounds are removed by the local background, the {\tt TBABS*APEC} model is adopted to fit the spectra. 
We use the AtomDB (version 3.0.8) database of atomic data and the solar abundance tables from \cite{2009ARA&A..47..481A}. For the hydrogen column density $N_{\rm H}$, we adopt the value $2.51 \times 10^{20}{\rm\ cm}^{-2}$ from the NHtot tool (\citealt{2013MNRAS.431..394W}). 
	
\section{results} 
	\label{sec:results_and_analysis}
\subsection{Global X-ray morphological properties}
	\label{sub:Global X-ray morphological properties}

The top right image of Fig.~\ref{fig:fig1} shows the 0.7 - 2 keV image of the snail subcluster J0157 from three \cha\ observations combined. The image is background-subtracted and exposure-corrected. 
The X-ray images are smoothed with a uniform Gaussian of 2.95 arcsec.
The image shows that the central part of subcluster J0157 has very bright X-ray emission, with the hot gas displaying a distinct spiral-shaped CF. 
The bright CF gas structure extends to about 260 kpc, with a tail trailing southward. A portion of the main cluster's hot gas structure is visible in the southwest, where the extended gas tail from the merging activity connects the subcluster to the main cluster, this connection is consistent with the \xmm\ observations (\citealt{Yang2026}).

The brightest cluster galaxy (BCG) at the center of J0157 hosts an active galactic nucleus (AGN). X-ray cavity-like features are visible around the BCG in the \cha\ image (Fig.~\ref{fig:fig1}). The cavities are spatially associated with the central AGN, suggesting that they have been inflated by AGN feedback. We do not attempt to quantify the cavity energetics in this work, but the presence of these cavities indicates that AGN activity may have affected the thermal and dynamical state of the intracluster gas in the central region of the subcluster.
Because the cavities are confined to the central region, their impact on the $\sim$260 kpc-scale spiral CF is expected to be limited. We then characterize the CF and its azimuthal variations by performing a detailed analysis in four sectors (Fig.~\ref{fig:region_sbp_direction}), each with distinct orientation relative to the subcluster’s motion.

\subsection{CF analysis in different sectors}

	\begin{figure*}
		\centering 
		\includegraphics[width=1\columnwidth]{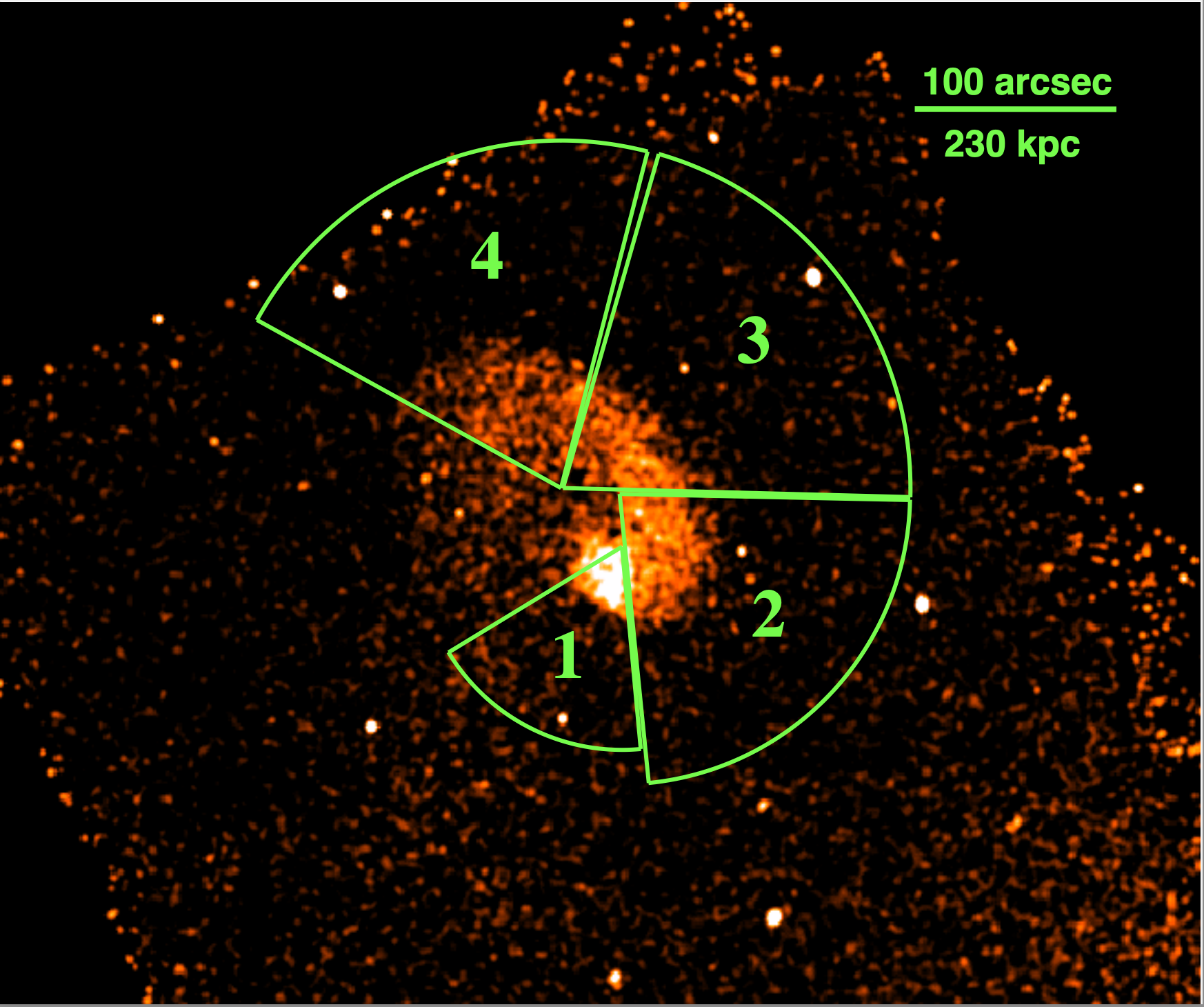}
        \includegraphics[width=0.998\columnwidth]{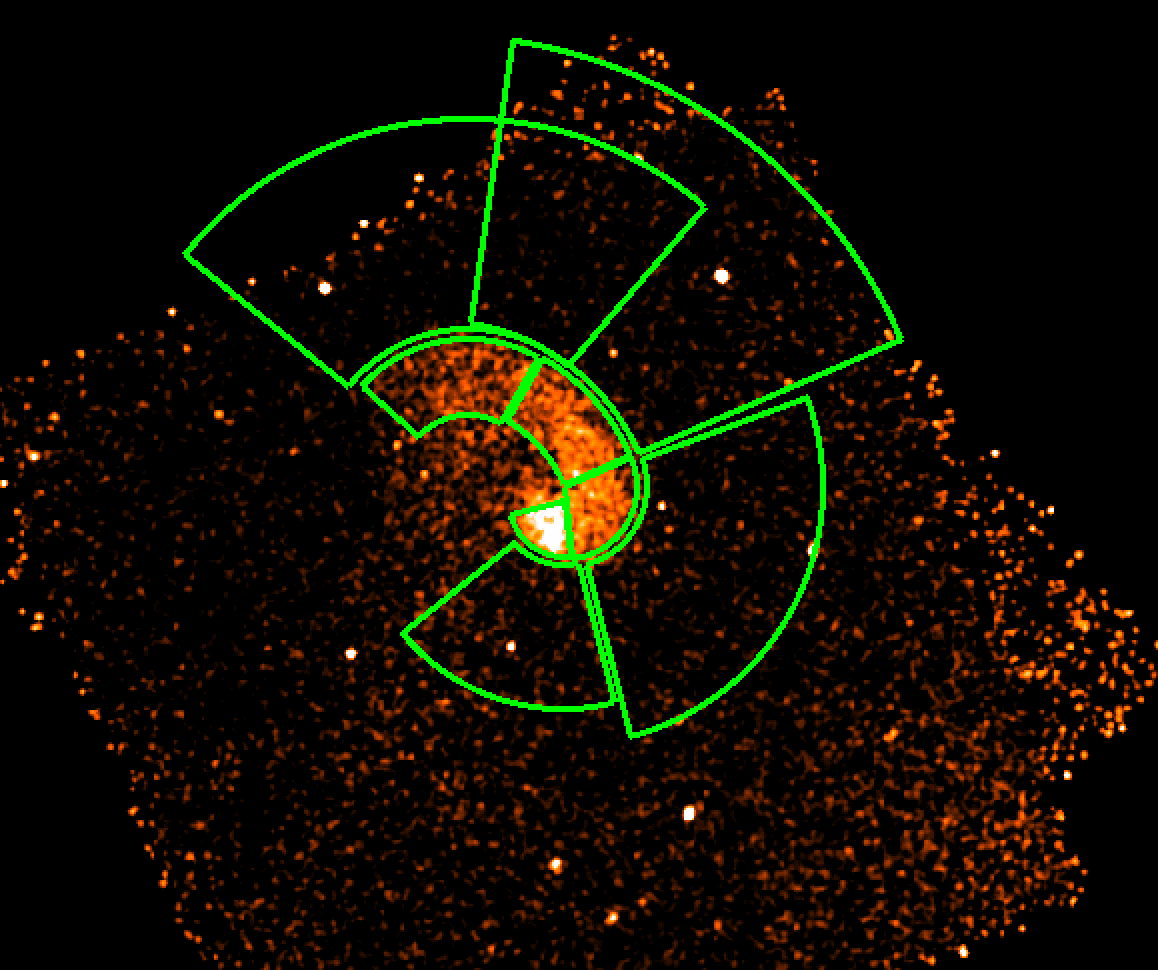}
		\caption{
		Merged \cha\ 0.7-2 keV image of J0157, with sectors used to extract radial profiles of the X-ray surface brightness and temperature in Fig.~\ref{fig:sbp_fit}. 
        Left: Extraction sectors in four regions for surface brightness.
        Right: Extraction sectors for temperature, because of the limited counts outside the front, mean temperatures are measured in larger regions.
		}
		\label{fig:region_sbp_direction}
	\end{figure*}
	
\begin{figure*}
		\centering
		\includegraphics[width=1\columnwidth]{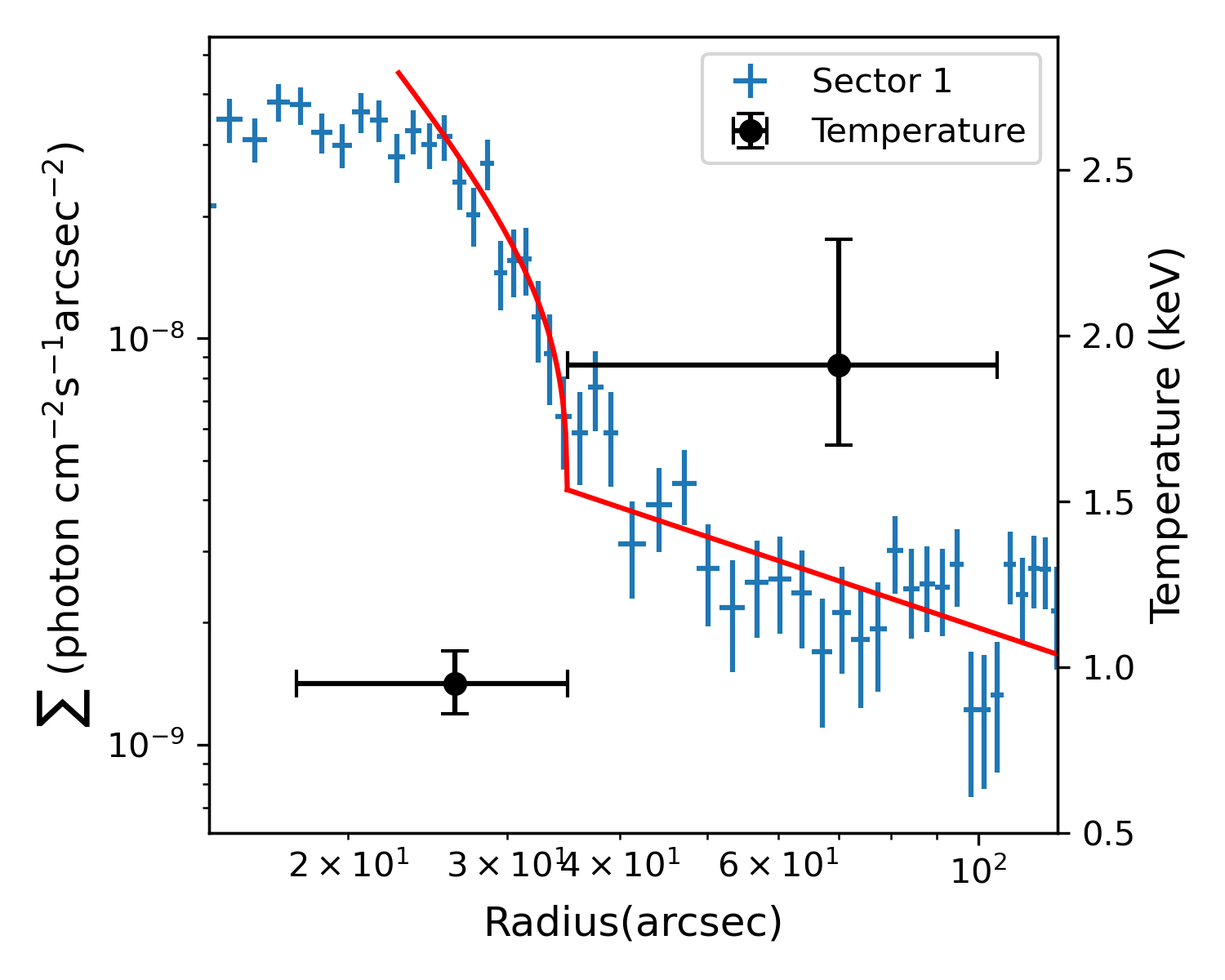}
		\includegraphics[width=1\columnwidth]{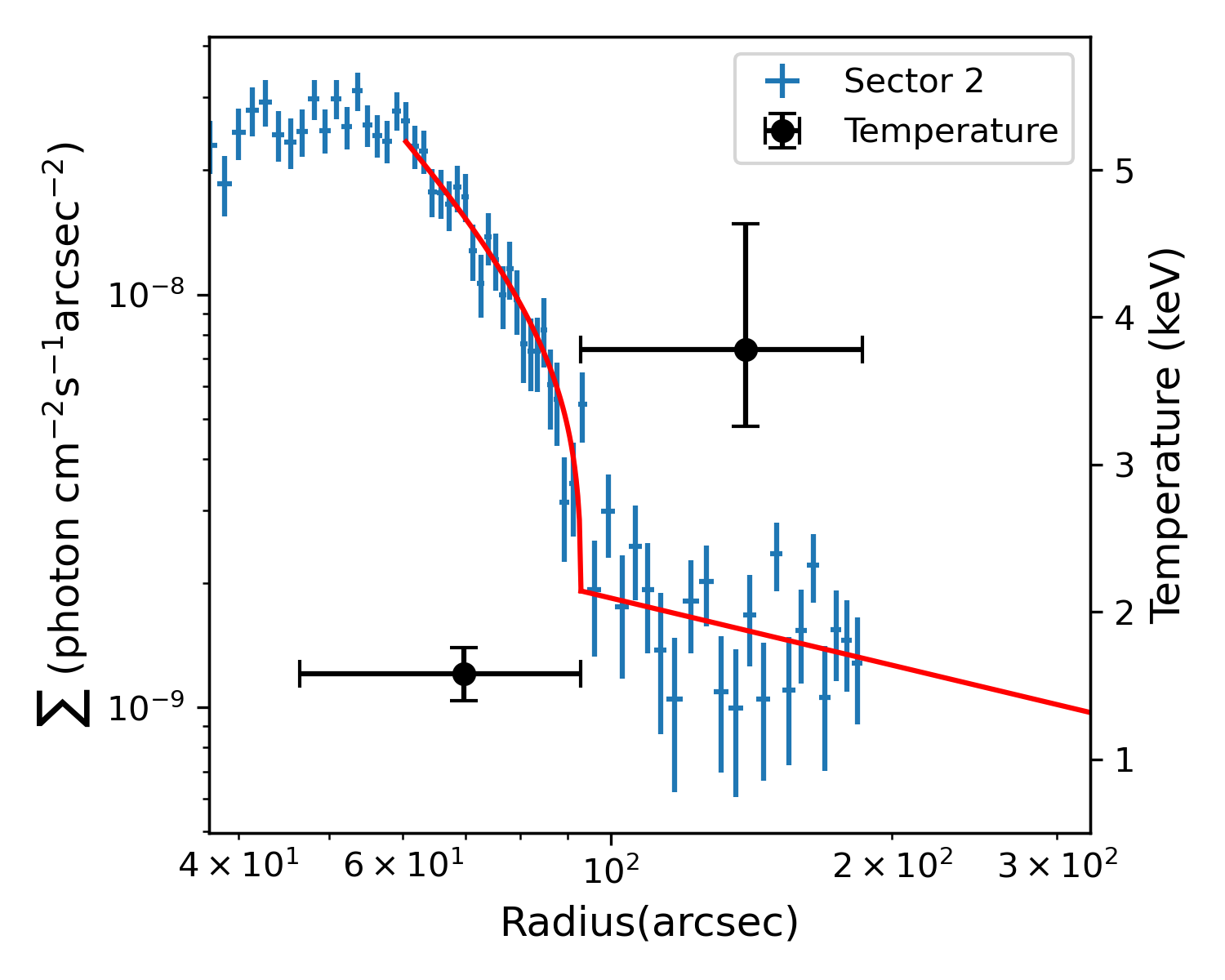}
		\includegraphics[width=1\columnwidth]{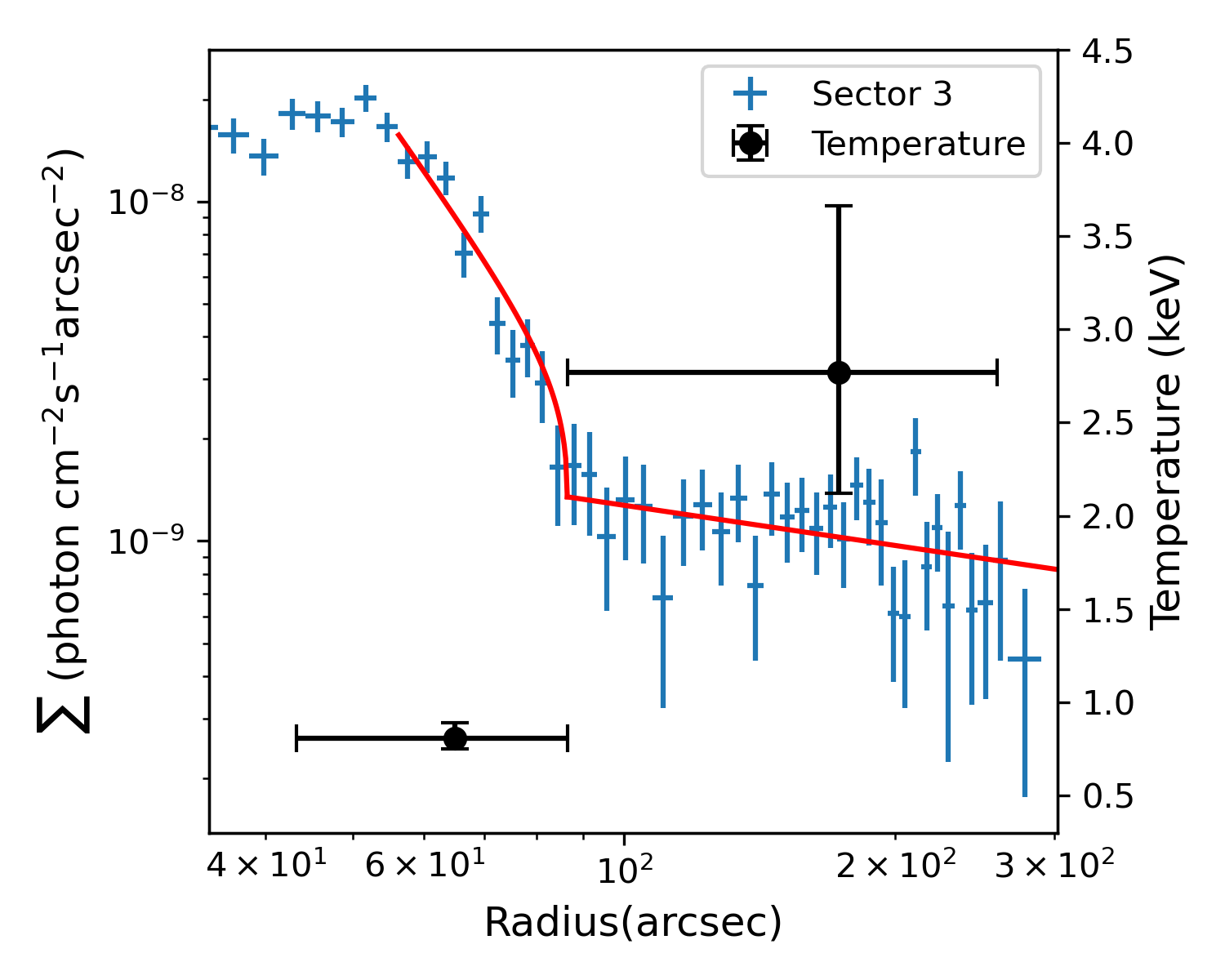}
		\includegraphics[width=1\columnwidth]{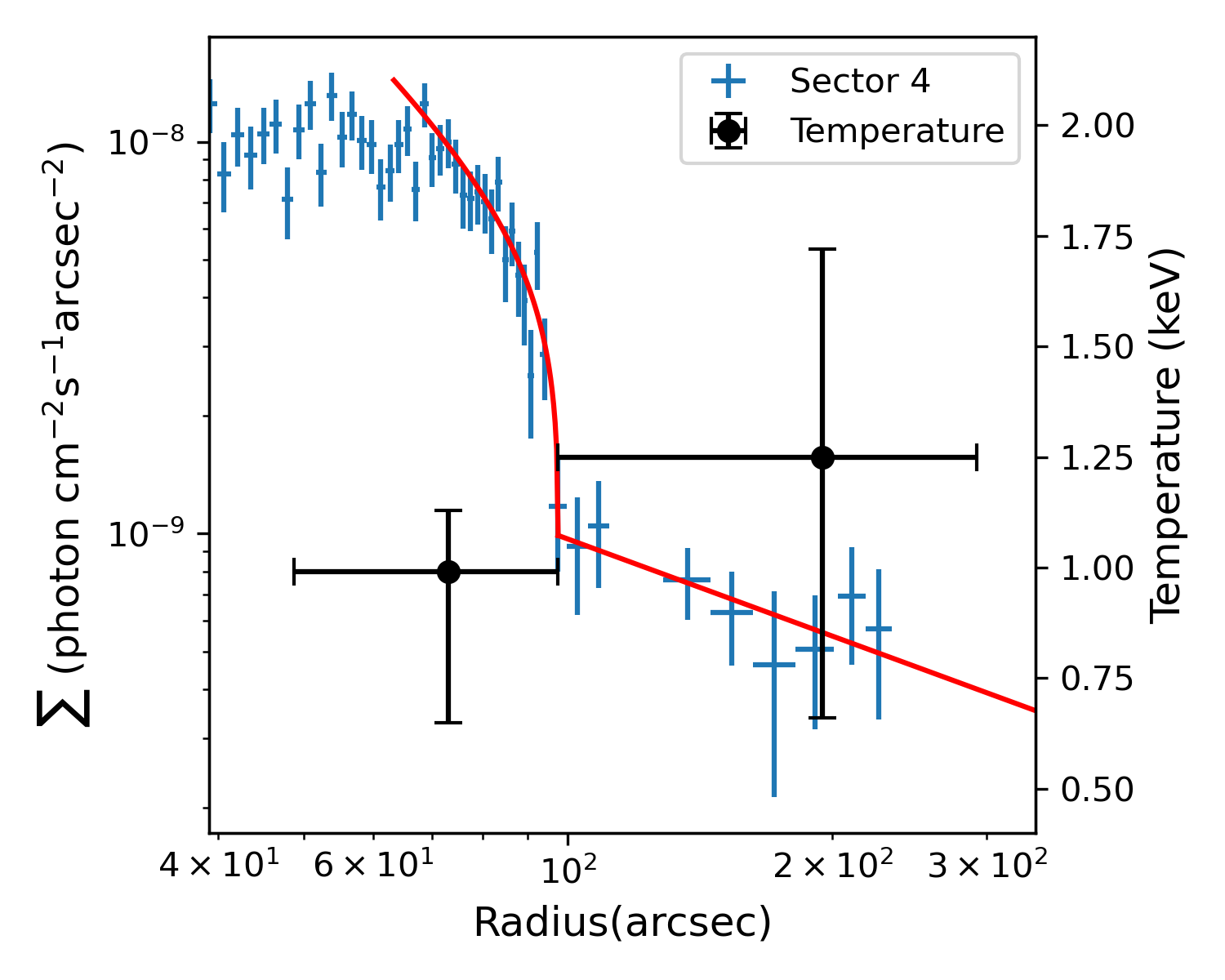}
		\caption{
		Surface brightness profiles of subcluster J0157 in the 0.7–2 keV band. The solid red line represents the best-fit model of the X-ray surface brightness profile. 
        The black points show X-ray temperatures on either side of the CFs.
		}
		\label{fig:sbp_fit}
	\end{figure*}

As shown in Fig.~\ref{fig:region_sbp_direction}, we define four sectors (1-4) at different positions along the CF and with different orientations relative to the subcluster motion. The X-ray surface brightness profiles (SBPs, in the 0.7–2 keV band) in these regions are extracted to further analyze the properties of the ICM plasma.
For sector 1, the SBP is extracted using circular annuli, with the horizontal axis representing the radial distance from the centers of the circles. For sectors 2-4, the profiles are extracted in elliptical annuli with a curvature approximately matched to that of the CF to accurately fit the density jump at its position. In these elliptical profiles, the radius along the horizontal axis corresponds to the semi-minor axis of the ellipse. The SBPs for all sectors are shown in Fig.~\ref{fig:sbp_fit}.

Following \cite{2016arXiv160607433S}, we assume the X-ray emissivity has an ellipsoidal discontinuity and follows a power-law in elliptical radius on each side of the front:
\begin{equation}
	\label{equa:emissivity}
	\epsilon(r)= \begin{cases}\epsilon_{i}\left(r / r_{\text {edge }}\right)^{-2 p_{i}} & r<r_{\text {edge }} \\ \epsilon_{o}\left(r / r_{\text {edge }}\right)^{-2 p_{o}} & r>r_{\text {edge .}}\end{cases}
\end{equation}
The X-ray surface brightness is given by
\begin{equation}
	I_{X}(r)=I_{\mathrm{in}}(r)+I_{\mathrm{out}}(r), 
\end{equation}
where $I_{\mathrm{in}}(r)$ and $I_{\mathrm{out}}(r)$ are given by
\begin{equation}
	\label{equa: Ii}
	I_{\mathrm{in}}(r)=I_{i} A^{-2 p_{i}+1} \begin{cases}1-I_{A^{2}}\left(p_{i}-\frac{1}{2}, \frac{1}{2}\right) & A^{2}<1 \\ 0 & A^{2} \geq 1\end{cases}, 
\end{equation}		
\begin{equation}
	\label{equa: Io}
	I_{\mathrm{out}}(r)=I_{o} A^{-2 p_{o}+1} \begin{cases}I_{A^{2}}\left(p_{o}-\frac{1}{2}, \frac{1}{2}\right) & A^{2}<1 \\ 1 & A^{2} \geq 1\end{cases} , 
\end{equation}
 where $A\equiv r/r_{\rm {edge}}$, $I_x(a,b)\equiv B_x(a,b)/B(a,b)$ is the normalized incomplete beta function, $B_x(a,b)\equiv \int_{0}^{x}t^{a-1}(1-t)^{b-1}dt$ is the incomplete beta function, $B(a,b)=\Gamma(a)\Gamma(b)/\Gamma(a+b)$ is the beta function, and $\Gamma(a)$ is the gamma function.
The $I_{o}$ and $I_{i}$ are the surface brightness outside and inside the edge respectively, and are normalized with
\begin{equation}
	I_{i}=R I_{o} \frac{B\left(p_{i}-\frac{1}{2}, \frac{1}{2}\right)}{B\left(p_{o}-\frac{1}{2}, \frac{1}{2}\right)},
\end{equation}
where $R \equiv \epsilon_{i} / \epsilon_{o}$ gives the jump in X-ray emissivity at the edge. 

Fig.~\ref{fig:sbp_fit} shows the fitted surface brightness profiles on both sides of the edge using Equations~(\ref{equa: Ii}) and (\ref{equa: Io}).
The emissivity jump values in these sectors are listed in Table~\ref{tab: results of CF}. 
The X-ray emissivity $\epsilon=\Lambda(T,Z)  n_e^2$ depends on gas temperature, metallicity, and density. 
The emissivity functions $\epsilon=\Lambda(T,Z)$ can be derived from the spectral fits for inner and outer regions of CFs used to derive the temperatures.
Then the gas density ratio across the CF can be derived from the emissivity jump \citep{2016arXiv160607433S}. The resulting density ratios for each direction are listed in Table~\ref{tab: results of CF},
which shows that the density ratio across the CF is approximately a factor of three in all sectors.

In addition, we perform temperature fitting for the gas on both the inner and outer sides of the CF in each sector. For the outer region, a larger extraction area was chosen to ensure sufficient photon statistics for spectral fitting. The extraction regions are illustrated in Fig.~\ref{fig:region_sbp_direction}.
As outlined in Section~\ref{sec: data analysis}, the gas on the inner side of the CF was fitted using a local background from the same CCD to exclude emission from the main cluster’s diffuse gas. For the outer side, the stowed background was adopted due to the dominant diffuse emission from the main cluster. The resulting temperatures are listed in the temperature column of Table~\ref{tab: results of CF}.
A clear temperature jump is observed across the CF, with variations in both the inner and outer temperatures as a function of direction relative to the CF.

	\begin{table*}
		\centering
		\caption{Properties of the CF in 4 different sectors.}
		\tabcolsep=0.2cm
		\begin{tabular}{lcccccc}
			\hline\hline 
			Sector  & Inner $kT$    & Outer $kT$   & Emissivity jump & Density ratio & Inner $n_e$  & Pressure ratio \\
           &    keV    & keV & & & $10^{-3}{\rm\ cm}^{-3}$  &  $P_{\text{ratio}}$  \\
			\hline
			 1 & $0.95_{-0.09}^{+0.10}$ & $1.91_{-0.24}^{+0.38}$   &  $7.8 \pm 4.6 $           & $2.8 \pm 0.8 $     &    7.4   & $1.5 \pm 0.6 $         \\
			 2 & $1.58_{-0.18}^{+0.18}$ & $3.78_{-0.52}^{+0.85}$   &  $13.3 \pm 8.8 $           & $3.7 \pm 1.2 $       & 4.2    & $1.6 \pm 0.8 $         \\
			 3 & $0.81_{-0.06}^{+0.08}$ & $2.77_{-0.65}^{+0.89}$   &  $10.9 \pm 5.5 $           & $3.3 \pm 0.8 $      & 2.9     & $1.0 \pm 0.5 $         \\
			 4 & $0.99_{-0.34}^{+0.14}$ & $1.25_{-0.59}^{+0.47}$   &  $13.2 \pm 6.1 $           & $3.6 \pm 0.8 $      & 2.2     & $3.0 \pm 1.5$          \\
			\hline
		\end{tabular}
		\label{tab: results of CF}
	\end{table*}
		
\subsection{Thermodynamic maps}

\begin{figure*}
		\centering
		\includegraphics[width=2\columnwidth]{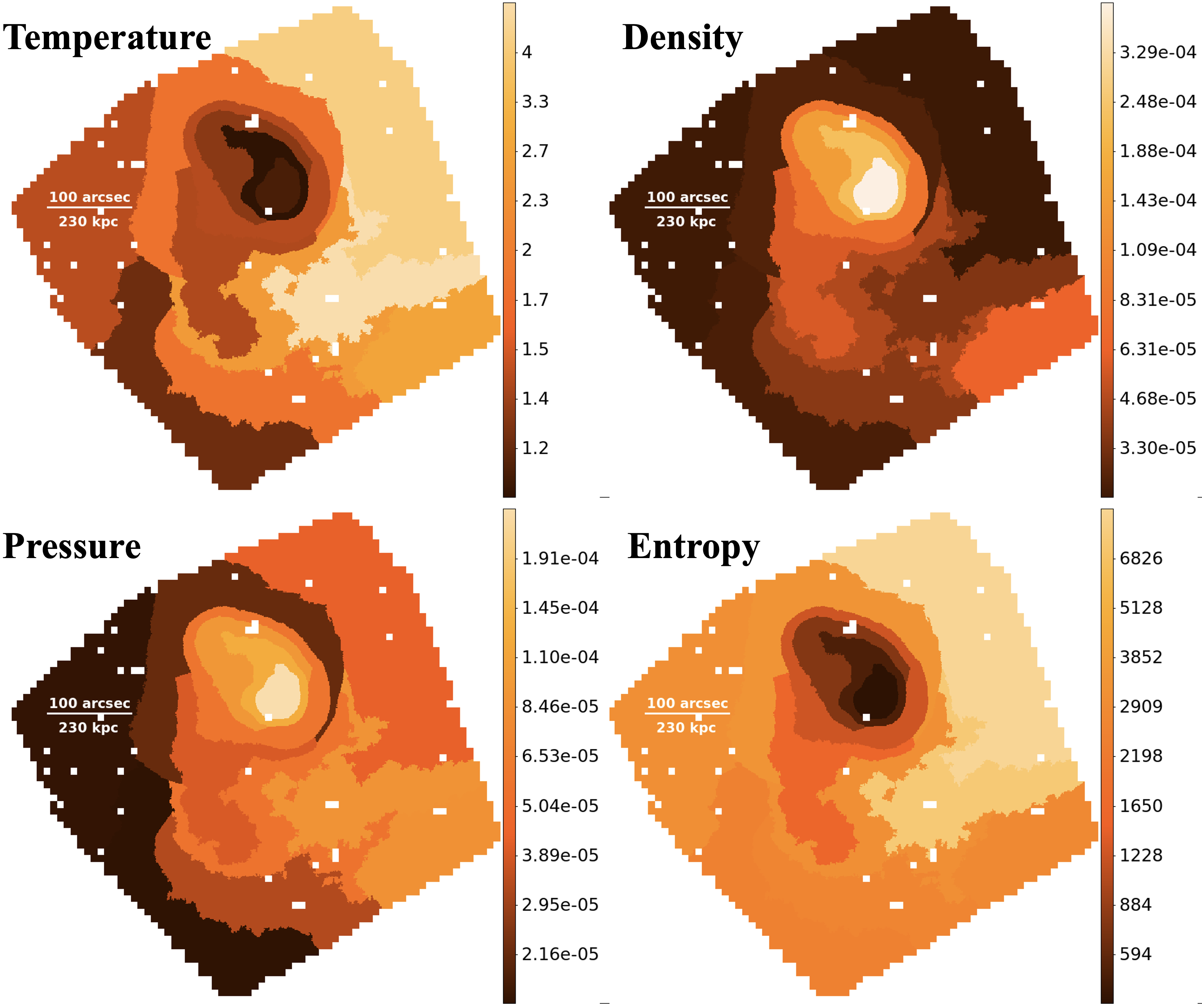}
		\caption{
		Projected \cha\ 2D maps of the ICM thermodynamic properties, generated by contour binning. 
		Top left: temperature.
		Top right: pseudo-density.
		Bottom left: pseudo-pressure.
		Bottom right: pseudo-entropy.
        }
		\label{fig:2D_map}
	\end{figure*}

To better understand the thermodynamics of the ICM around the subcluster J0157, 
we partition the \cha\ ACIS X-ray image using the ``Contour binning" algorithm \citep{2006MNRAS.371..829S}. 
This technique groups pixels by surface brightness so that each spectral extraction region contains sufficient X-ray photon counts.
After masking the point sources in the X-ray image, we set the signal-to-noise ratio parameter to 40, which requires an effective number of photons per partition of $\sim1600$ after subtracting the number of photons from the background.
In addition, we set the constraint geometry parameter value to 2 to ensure that the shape of each partition is not too elongated. The final regions used to extract the spectra are shown in Fig.~\ref{fig:2D_map}.
We use these regions to extract the spectra and response files. Then we fit the temperature of the gas according to the method described in detail in Section~\ref{sec: data analysis}, and the results are shown in the top-left panel of Fig.~\ref{fig:2D_map}.

We can calculate the mean gas density in Table~\ref{tab: results of CF} and the projected gas density from the parameters of the thermal model used in fitting the spectra \citep[e.g.][]{2018MNRAS.476.5591B,Ge2021}.
The normalization parameter in the thermal gas model is given by:
\begin{equation}
\mathcal{N}=\frac{10^{-14}}{4 \pi\left[D_{\mathrm{A}}(1+z)\right]^{2}} \int n_{\mathrm{e}} n_{\mathrm{H}} \mathrm{d} V \quad\left(\mathrm{~cm}^{-5}\right),
\end{equation}
where $D_{\mathrm{A}}$ is the angular diameter distance to the observed source, and $n_{\mathrm{e}}$ and $n_{\mathrm{H}}$ are the densities of electron and hydrogen, respectively.
The projected normalization per unit area is
\begin{equation}
N_{\rm proj}=\mathcal{N} / A \quad\left(\mathrm{~cm}^{-5} \operatorname{arcsec}^{-2}\right),
\end{equation}
where A represents the angular area of each partition, and $N_{\rm proj}$ is proportional to the square of the electron density integrated along the line of sight.
The projected gas pressure is defined as
\begin{equation}
P=k T(N_{\rm proj})^{1 / 2} \quad\left(\mathrm{keV}\ \mathrm{cm}^{-5 / 2} \operatorname{arcsec}^{-1}\right).
\end{equation}
The projected gas entropy is defined as
\begin{equation}
K=k T(N_{\rm proj})^{-1 / 3} \quad\left(\mathrm{keV}\ \mathrm{cm}^{5 / 3} \operatorname{arcsec}^{2 / 3}\right).
\end{equation}
Using these definitions, we derived the {projected} density, pressure, and entropy values for each partitioned region. The results are presented in Fig.~\ref{fig:2D_map}, whose four panels display the projected thermodynamic properties: temperature (top left), pseudo-density (top right), pseudo-pressure (bottom left), and pseudo-entropy (bottom right).
It is important to note that these maps represent line-of-sight projected quantities rather than the true three-dimensional distributions of these physical parameters.
Nevertheless, these projected maps effectively illustrate the relative distributions of thermodynamic properties across the system (e.g. \citealt{Hu_2021,Sanders2022}).

The temperature map (Fig.~\ref{fig:2D_map} top left panel) exhibits a distinct temperature jump across the CF in the subcluster,  quantitatively consistent with our spectral analysis in multiple sectors (Table~\ref{tab: results of CF}). 
The extended curved tail-like structure remains markedly cooler than the surrounding ambient gas, indicating that this gas originated from the subcluster's CC and was subsequently stripped during the merger process.
The density map (Fig.~\ref{fig:2D_map} top right panel) reveals a well-defined CF and a prominent stripped gas tail, with a pronounced density discontinuity evident across the CF interface.
Compared to the sharp contrasts in the density and temperature distributions, the pressure variation (Fig.~\ref{fig:2D_map} bottom left panel) across the CF interface is relatively modest.
A low-pressure band appears along the outer edge of the CF. This feature may be explained by the ram pressure of the ambient gas moving relative to the CF. Because the outer gas flows around the CF, its thermal pressure is reduced relative to that on the inner side by the ram pressure contribution (see detailed discussion in Section~\ref{sec:gas_dynamics}).
The entropy distribution (Fig.~\ref{fig:2D_map} bottom right panel) traces the thermodynamic structure of the system. It closely mirrors the temperature structure.
The subcluster's ICM maintains systematically lower entropy values within the CF boundary, while the main cluster's ICM exhibits distinctly higher entropy levels.
Collectively, the projected thermodynamic maps reveal systematic variations across the CF interface between the subcluster and main cluster.

\section{discussion}
\subsection{Merger dynamics} 
	\label{sec:gas_dynamics}	

\begin{figure*}
		\centering 
		\includegraphics[angle=0,width=0.98\textwidth]{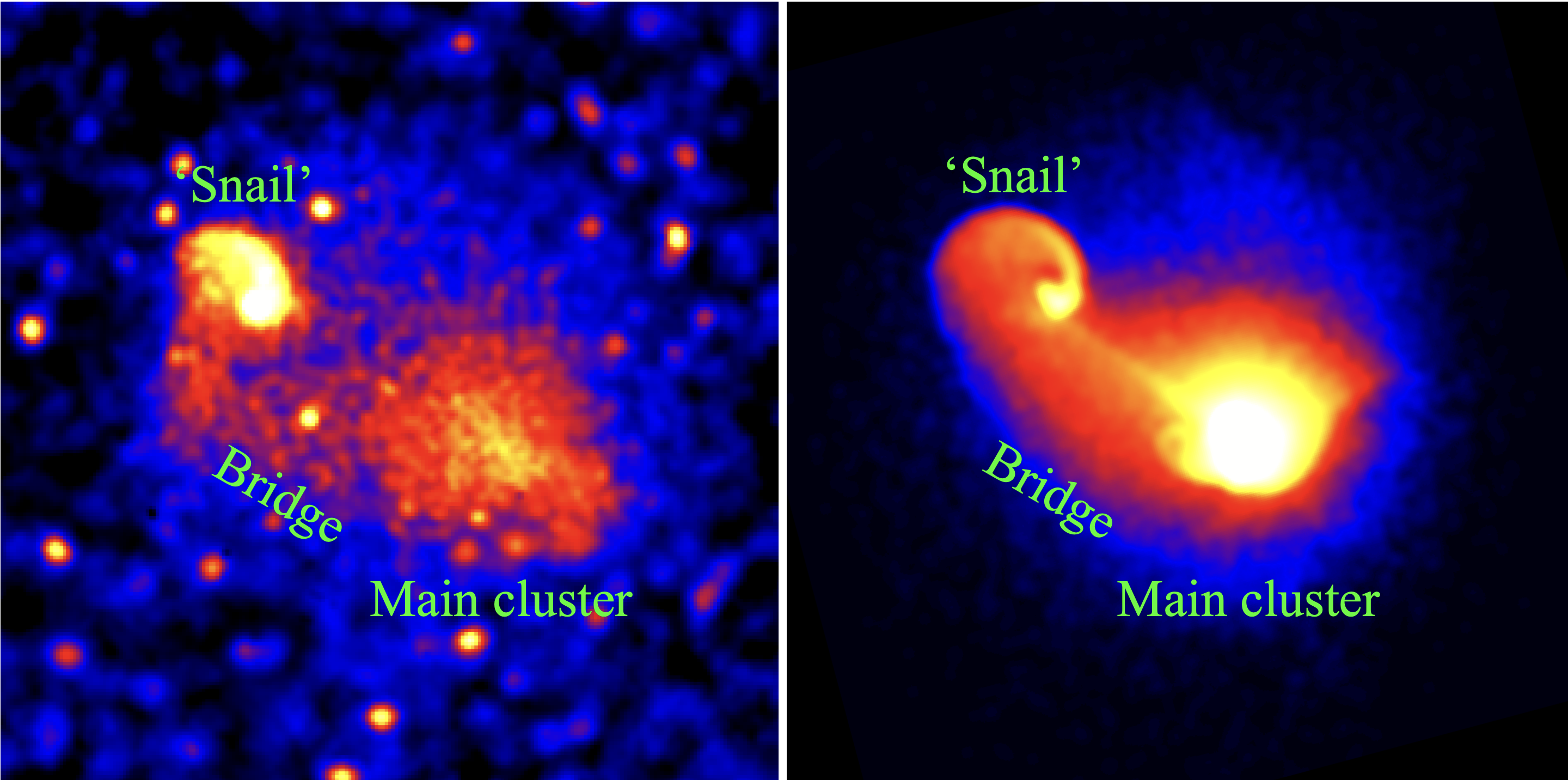}
		\caption{Comparison between the \xmm\ 0.5-2 keV image and a mock X-ray image from simulation. The simulation presents a merger between two clusters with a mass ratio of 3, an initial impact parameter of 1000 kpc, and a merger time of 2.3 Gyr from the beginning of the simulation \citep{ZuHone2011,ZuHone2018}. The labels highlight similar substructures between observation and simulation, including the main cluster, the `snail' subcluster, and the gas bridge between them.}
		\label{fig:sim}
	\end{figure*}

Fig.~\ref{fig:sim} compares the \xmm\ image with a mock X-ray image extracted from a simulation of a merger between two clusters with a mass ratio of 3, an initial impact parameter of 1000 kpc, and an elapsed time of 2.3 Gyr since the starting point of the simulation \citep{ZuHone2011, ZuHone2018}. The simulated system reproduces the key structural components seen in the data: a massive, relatively relaxed main cluster; a smaller subcluster displaying a well-developed spiral pattern analogous to the observed `snail'; and a diffuse bridge of gas connecting the two components \citep{Yang2026}. This close morphological correspondence supports the interpretation that subcluster J0157 is not falling into the main cluster along a purely radial trajectory, but rather on an orbit with a substantial non-zero impact parameter. Such an impact parameter is required to impart sufficient angular momentum to the displaced gas to generate a large-scale, coherent spiral pattern. The fact that the spiral front in J0157 remains coherent out to $\sim 260$ kpc is therefore naturally accommodated by this merger scenario.
In the following subsections, we use the pressure jumps measured across the CF in different sectors to derive more quantitative constraints on the merger velocity and geometry implied by this picture.    

The bulk motion of the subcluster can be constrained by the pressure balance across the CF. Assuming that the interface is in instantaneous equilibrium between the thermal pressure of the cooler gas inside the front (\(p_c\)) and the sum of the thermal pressure and ram pressure of the exterior hotter gas (\(p_h + \rho_h v^2\)), we have
\begin{equation}
p_c = p_h + \rho_h v^2,
\end{equation}
where \(\rho_h\) is the mass density of the gas outside the front and \(v\) is the relative velocity of the subcluster with respect to the ambient medium \citep{Markevitch2000}. 
The ideal gas law gives \(p_h = n_{\text{tot},h} k T_h = (\rho_h / \mu m_p) k T_h\). Substituting into the pressure balance equation yields
\begin{equation}
\rho_h v^2 = p_h\left(\frac{p_c}{p_h}-1\right) = p_h (P_{\text{ratio}}-1),
\end{equation}
\begin{equation}
v^2 = \frac{p_h}{\rho_h} (P_{\text{ratio}}-1) = \frac{k T_h}{\mu m_p} (P_{\text{ratio}}-1).
\end{equation}
Hence the velocity is
\begin{equation}
v = \sqrt{(P_{\text{ratio}}-1) \cdot \frac{k T_h}{\mu m_p}}.
\end{equation}
Using \(m_p = 1.67\times10^{-27}\) kg, and \(\mu=0.6\), we obtain the numerical expression
\begin{equation}
v = \sqrt{(P_{\text{ratio}}-1)\; T_h\;(\text{keV})}\;\times\;400\;\text{km s}^{-1}.
\end{equation}
Here \(P_{\text{ratio}} = p_c/p_h\) is the thermal pressure ratio across the front, and \(T_h\) is the gas temperature outside the front in keV. Using the measured pressure ratios and outer temperatures from Table~\ref{tab: results of CF}, we can compute the velocity in each sector.

The largest inferred velocity is found in sector~4, reaching \(630\) km s\(^{-1}\), while sector~3 shows a pressure ratio consistent with unity, implying negligible normal velocity. 
This is consistent with sector 3 being oriented nearly tangentially to the inferred motion. Sectors~1 and~2 exhibit intermediate velocities of \(390\) and \(600\) km s\(^{-1}\). 
The azimuthal variation in the inferred velocities may reflect the changing orientation of the CF relative to the gas motion and could be related to the orbital and tidal dynamics of the stripped subcluster gas. Together with the near-unity pressure ratio in sector 3 and the higher ratio in sector 4, this pattern is consistent with the subcluster moving toward the northeast, broadly in agreement with the merger geometry suggested by the simulations \citep{ZuHone2011,ZuHone2018}.
This picture is independently corroborated by \cite{Yang2026}, who combined Subaru/HSC weak-lensing shape measurements with an analytic merger-dynamics formula incorporating dynamical friction \citep{Okabe2026} to constrain the merger geometry, and showed that the observed spiral pattern arises from tidally induced rotation following core passage. Their dynamical modeling yields an infall velocity of \(\sim 1000\ \mathrm{km\ s^{-1}}\), an impact parameter of \(\sim 0.9\ \mathrm{Mpc}\), and a current subcluster velocity of \(\sim 1135\ \mathrm{km\ s^{-1}}\). 
The hot gas velocity inferred from the pressure balance ($\sim 630$ km s$^{-1}$) is lower than the orbital velocity of the subcluster’s dark matter halo ($\sim 1135$ km s$^{-1}$). The two quantities need not agree: the former is the speed of the gas at the front relative to the ambient ICM, whereas the latter is the bulk velocity of the dark matter halo.

\subsection{Kelvin–Helmholtz instabilities}
\label{sec:KHI}

\begin{figure*}
		\centering 
		\includegraphics[angle=0,width=0.98\textwidth]{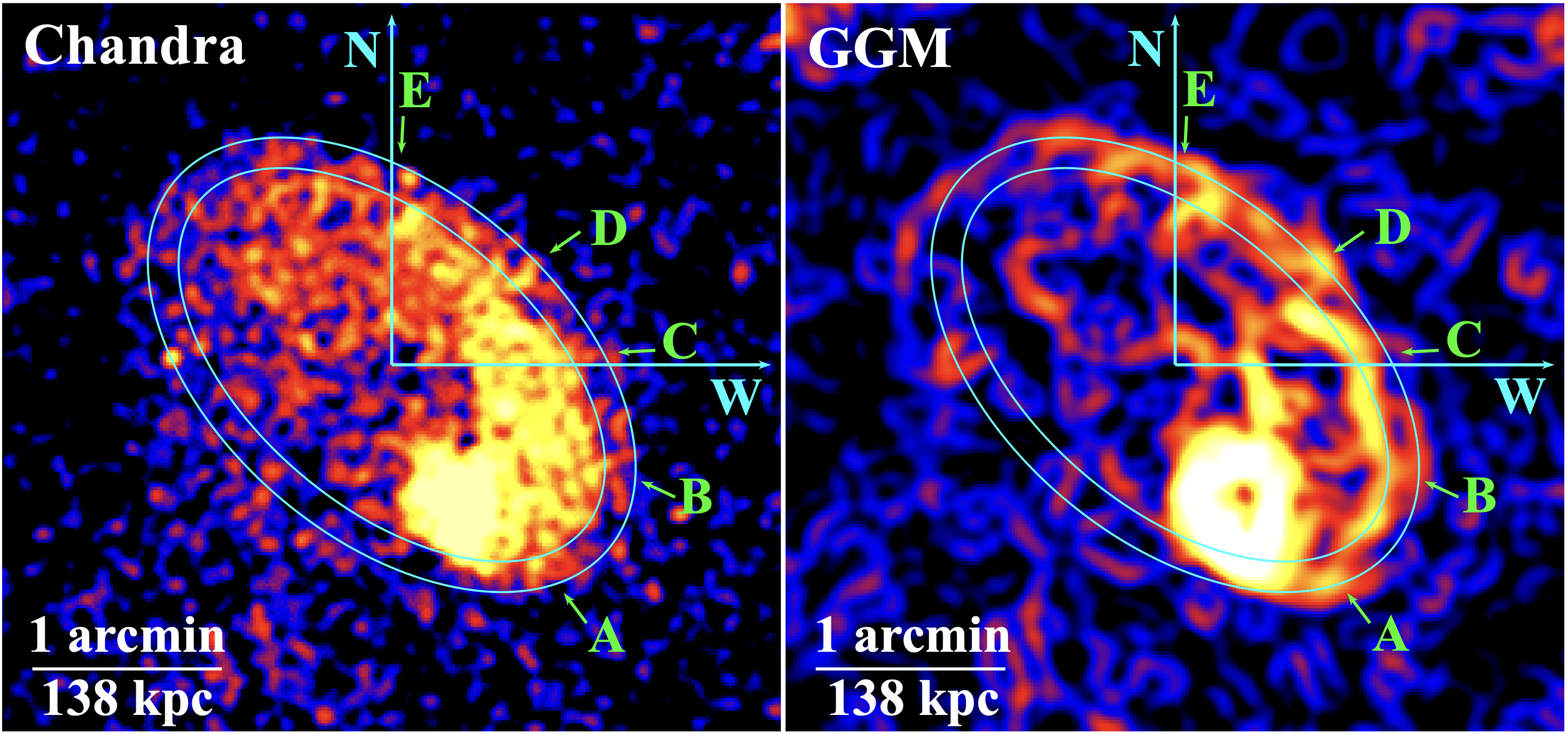}
        \includegraphics[angle=0,width=0.98\textwidth]{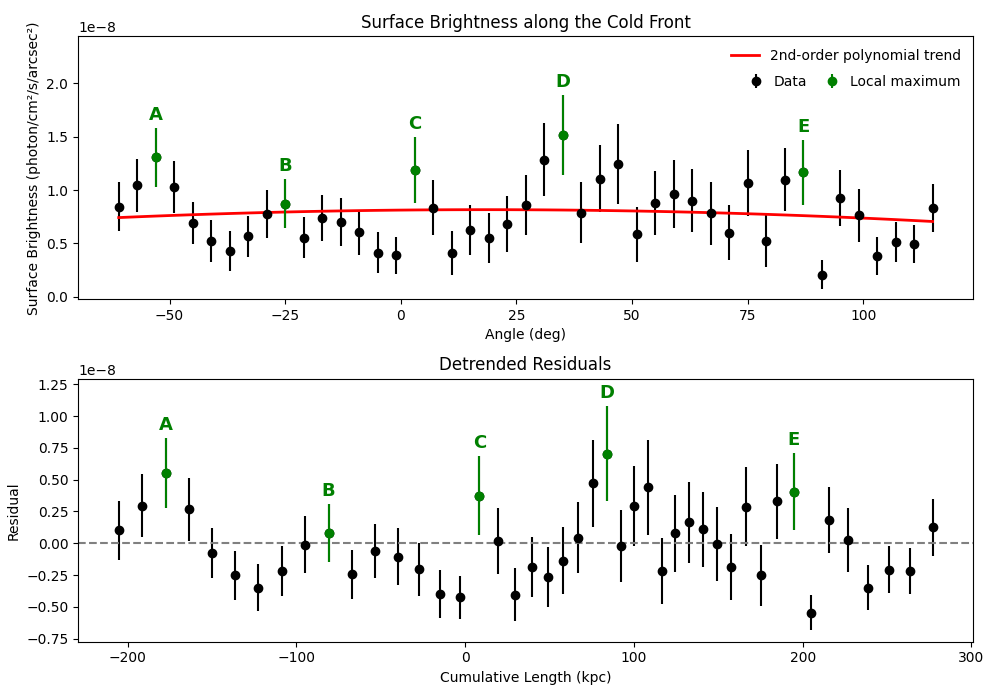}
		\caption{Top: \cha\ and GGM image binned by a factor of 2 (1 pixel is 0.984 arcsec) and a smooth width $\sigma = 4$ pixels, A-E mark the local maximum as the potential KH rolls. West and north correspond to 0$^\circ$ and 90$^\circ$, respectively. The azimuthal SBPs are extracted within the cyan elliptical annulus from -63$^\circ$ to 117$^\circ$ in 4$^\circ$ bins. Middle: The azimuthal SBPs along the CF in degrees. Bottom: the residual surface brightness after subtracting the 2nd-order polynomial trend with the $x$-axis converted to kpc.
}
		\label{fig:arc}
	\end{figure*}

As contact discontinuities sustaining velocity shear, CFs are expected to be distorted by KH instabilities, unless their growth is damped by the ICM viscosity or suppressed by magnetic fields aligned with the interface \citep[e.g.][]{Markevitch2007,Zuhone2016}. 
The pressure analysis in Section~\ref{sec:gas_dynamics} implies relative velocities of several hundred km s$^{-1}$ between the stripped gas and the ambient medium, making the CF of J0157 a promising site for KH instabilities.
While the spiral CF in J0157 appears smooth in the raw \cha\ image (Fig.~\ref{fig:fig1}), a closer examination reveals possible evidence for corrugation of the interface. We applied the GGM filtering technique \citep{Sanders2016} to the merged 0.7–2 keV image to enhance surface brightness edges. The resulting GGM map (Fig.~\ref{fig:arc}, top right) reveals a series of wave-like, finger-shaped protrusions along the front. 

To quantify this apparent corrugation, we extracted the azimuthal SBP along the front within an elliptical annulus whose curvature approximately matches that of the CF, from a azimuthal angle of $-63^\circ$ to $117^\circ$ (west and north correspond to $0^\circ$ and $90^\circ$, respectively) in bins of $4^\circ$. The profile was detrended by subtracting a 2nd-order polynomial fit to the large-scale azimuthal variation (Fig.~\ref{fig:arc}, middle and bottom panels). The residual profile reveals five discrete local maxima, labeled A–E in Fig.~\ref{fig:arc}. The positions of A–E coincide, within the resolution of the SBP extraction, with individual wave-like features identified independently in the GGM map, providing two complementary views of evidence for the same underlying structures. We interpret A–E as candidate KH rolls or billows, distorting the CF, analogous to structures reported in other well-studied CFs (e.g. \citealt{Ichinohe2017,Wang2018}).

Converting the azimuthal positions of A–E into a physical arc length along the elliptical front, the candidate rolls are separated by 75–111 kpc (mean characteristic scale of $\sim$93 kpc), spanning a total of $\sim$370 kpc from A to E. 
An alternative method for estimating the characteristic scale of the KH rolls is to use the height of the rolls, which approximately corresponds to the width (10 arcsec) of the elliptical annulus. The characteristic KH scale is expected to be several times this height \citep{Roediger2013}, adopting a factor of 4 gives a characteristic scale of $\sim$92 kpc.
Thus, the characteristic scale from the quasi-periodic structure defined by the local maxima in the azimuthal surface-brightness profile is consistent with the one estimated from the KH roll height.
These scales are comparable to those reported in other clusters where KH instabilities have been claimed: A3667 (20-80 kpc; \citealt{Ichinohe2017}), 
Perseus cluster ($\sim$50 kpc; \citealt{Walker2017}), 
A2142 ($\sim$55 kpc; \citealt{Wang2018}), 
A2319 ($\sim$100 kpc; \citealt{Ichinohe2021}). 
The presence of these KH rolls in J0157 indicates that the effective viscosity and/or the stabilizing magnetic layer are not sufficient to suppress all KH modes along the front.

\subsection{Constraints on the microphysics}

The candidate KH rolls identified in Section~\ref{sec:KHI} provide a valuable probe of the microphysical properties of the ICM. In particular, the presence of these instabilities indicates that the shear layer is not fully stabilized by either the effective viscosity or a magnetic layer aligned with the CF. This allows us to place an upper limit on the effective viscosity, and indirectly to constrain the strength of the magnetic field that can be sustained at the interface.

First, we place an upper limit on the effective viscosity of the ICM near J0157 from the observed KH rolls. 
The KH instability grows when the Reynolds number (Re) of the flow is above a critical value (e.g. \citealt{Chandrasekhar1961,Roediger2013}): 
\begin{equation}
	\operatorname{Re}=\frac{L U}{\nu}>\operatorname{Re}_{\text {crit }},
\end{equation}
where $L$ is the characteristic perturbation scale, $U$ is the shear velocity, and $\nu$ is the kinematic viscosity, which scales as $\nu \propto T^{5/2} n^{-1}$ in an unmagnetized plasma \citep{Spitzer1956}. Thus, the Reynolds number for a Spitzer-like viscosity \citep{Roediger2013} is
\begin{equation}
\begin{split}
\mathrm{Re}=10\,f_\mu^{-1}
&\left(\frac{U}{400\,\mathrm{km\,s^{-1}}}\right)
\left(\frac{L}{10\,\mathrm{kpc}}\right) \\
&\times
\left(\frac{n_e}{10^{-3}\,\mathrm{cm^{-3}}}\right)
\left(\frac{kT_{\mathrm{ICM}}}{2.4\,\mathrm{keV}}\right)^{-5/2},
\end{split}
\end{equation}
where \(f_\mu=\mu/\mu_{\rm Sp}\) is the viscosity suppression factor relative to the Spitzer value. 
Due to the strong temperature dependence of Spitzer viscosity, the Reynolds numbers in the cold and hot layers of a shear flow differ significantly, scaling as $D_\rho^{7/2}$, where $D_\rho=\rho_c/\rho_h$ is the density ratio. Consequently, only the viscosity in the hotter layer is sufficient to suppress the instability \citep{Roediger2013}. Therefore, we use the ICM properties in the hot layer (outside the CF interface) to estimate the viscosity in sector 3, where most candidate KH rolls are found.
We then adopt shear velocity \(U=630\ \mathrm{km\ s^{-1}}\) from the pressure-balance estimate in Section~\ref{sec:gas_dynamics}, \(L\sim90\ \mathrm{kpc}\) as the characteristic KH roll scale inferred in Section~\ref{sec:KHI}, \(n_e=0.9\times10^{-3}\ \mathrm{cm^{-3}}\) and \(kT=2.8\ \mathrm{keV}\) as representative values for the gas outside the CF (Table~\ref{tab: results of CF}).

The critical Reynolds number above which the KH instability grows is given by \citep{Roediger2013,Ichinohe2017} \(\mathrm{Re}_{\rm crit}\approx64\sqrt{\Delta}\), where $\Delta=\left(\rho_{c}+\rho_{h}\right)^{2} / \rho_{c} \rho_{h}$ is the density contrast parameter.
For our measured density ratio  $D_\rho= \rho_c/\rho_h\approx 3$, this yields ${\rm Re_{crit}}\approx 148$. Requiring \(\mathrm{Re}>\mathrm{Re}_{\rm crit}\) for the observed KH rolls to develop yields $f_\mu \lesssim
0.59$.
Thus, the effective dynamic viscosity is limited to \(\mu\lesssim0.59\,\mu_{\rm Sp}\). At \(kT=2.8\ \mathrm{keV}\), the Spitzer dynamic viscosity is
\begin{equation}
\begin{split}
\mu_{\mathrm{Sp}} &\approx 1.0\times 10^{3}\,
\frac{(kT/\mathrm{keV})^{5/2}}{\ln\Lambda_i}
\ \mathrm{g\,cm^{-1}\,s^{-1}} \\
&\approx 3.3\times 10^{2}
\ \mathrm{g\,cm^{-1}\,s^{-1}},
\end{split}
\end{equation}
where the Coulomb logarithm is $\ln\Lambda_i\approx40$,
so that \(\mu\lesssim1.9\times10^{2}\ \mathrm{g\ cm^{-1}\ s^{-1}}\). 
This upper limit indicates that the ICM viscosity in J0157 is at most of the order of the Spitzer value, although substantially lower viscosities are not excluded. 
It is instructive to compare the viscosity suppression factor inferred for J0157 with constraints obtained from KH rolls in other galaxy clusters, e.g., Abell 3667 ($\lesssim 0.05$ Spitzer; \citealt{Ichinohe2017}),
Virgo Cluster ($\sim 0.1$ Spitzer; \citealt{Roediger2013a}),
Abell 2142 ($\lesssim 0.2$ Spitzer; \citealt{Wang2018}),
Abell 2319 ($\lesssim 0.4$ Spitzer; \citealt{Ichinohe2021}).

Second, the same KH rolls also imply that any magnetic layer at the front is not strong enough to suppress all unstable modes, which provides an indirect upper limit on the magnetic field strength at the interface. This limit depends on the time available for the KH instability to grow before the interface is significantly modified by the merger dynamics. \cite{Yang2026} showed that the spiral structure and the CF in J0157 are likely shaped by tidally induced rotation, which stretches and distorts the interface on a characteristic timescale \(t_{\rm tidal}\). For the KH rolls to be observable, the instability must grow sufficiently fast compared to this tidal timescale. We therefore require that the KH growth timescale be shorter than the tidal timescale, i.e.\ \(\gamma_{\rm KH}^{-1}<t_{\rm tidal}\), where \(\gamma_{\rm KH}\) is the KH growth rate. This condition provides an upper limit on the magnetic field strength, because a stronger field reduces \(\gamma_{\rm KH}\) and would prevent the rolls from growing within \(t_{\rm tidal}\).

The parallel-field KH dispersion relation is
\begin{equation}
\gamma_{\rm KH}=k\left[
U^2\frac{\rho_c\rho_h}{(\rho_c+\rho_h)^2}
-\frac{B^2}{2\pi(\rho_c+\rho_h)}
\right]^{1/2},
\end{equation}
where $k=2\pi/\lambda$ and $\lambda$ is the characteristic scale.
Requiring \(\gamma_{\rm KH}^{-1}<t_{\rm tidal}\) yields
\begin{equation}
B^2 < 2\pi(\rho_c+\rho_h)
\left[
U^2\frac{\rho_c\rho_h}{(\rho_c+\rho_h)^2}
-\frac{\lambda^2}{4\pi^2 t_{\rm tidal}^2}
\right].
\end{equation}
Using the gas parameters for sector~3, gas density $\rho=\mu m_p n_{\mathrm{tot}} = 1.93 \mu m_p n_e$ with \(n_e = 2.9\times10^{-3}\ \mathrm{cm^{-3}}\) inside the CF, density ratio \(D_\rho\sim3\), shear velocity \(U=630\ \mathrm{km\ s^{-1}}\) from the pressure-balance estimate in Section~\ref{sec:gas_dynamics}, \(\lambda\sim90\ \mathrm{kpc}\) as the characteristic KH roll scale inferred in Section~\ref{sec:KHI}, and \(t_{\rm tidal}=0.7\ \mathrm{Gyr}\) from \citealt{Yang2026}, we estimate this upper limit as 
$B \lesssim 5.9\ \mu\mathrm{G}.$
This upper limit is consistent with the presence of the observed KH rolls, which indicate that the magnetic layer at the front is not strong enough to suppress all unstable modes. 
The magnetic field of J0157 can also be compared with other CFs, e.g., 
A3667 ($\sim 6.8\ \mu$G; \citealt{Omiya2026}),
NGC 7618 \& UGC 12491 group ($< 2\ \mu$G; \citealt{Roediger2012}),
A496 ($< 10\ \mu$G; \citealt{Roediger2012a496}), 
NGC 1404 ($< 5\ \mu$G; \citealt{Su_2017}).

\section{Conclusions}

We have presented a detailed analysis of the tidally induced CF in the `snail' subcluster J0157, based on 130 ks of deep \cha\ observations. Our main results are summarized as follows:

\begin{itemize}
\item The subcluster J0157 exhibits an exceptionally large and well-defined spiral CF, extending to $\sim 260$ kpc from the subcluster center, with an associated tail traced to at least 700 kpc. This makes it one of the largest such features identified to date, and it is directly visible in the raw image without requiring subtraction of an azimuthally averaged model, unlike most classical sloshing CFs.
Surface brightness fitting across four sectors yields a consistent density jump of a factor of $\sim 3$ across the front in every direction, accompanied by significant temperature discontinuities. Projected thermodynamic maps (temperature, density, pressure, and entropy), constructed via contour binning, corroborate this picture and reveal a coherent, cool, low-entropy structure tracing the stripped subcluster gas.

\item By applying pressure balance across the front in each sector, we derive relative velocities between the stripped subcluster gas and the ambient medium ranging from near-zero (sector 3, where the front runs tangential to the motion) up to $\sim 630$ km s$^{-1}$ (sector 4). This azimuthal pattern is consistent with the subcluster moving toward the northeast. A morphology-comparison simulation with a mass ratio of 3 and an impact parameter of 1000 kpc qualitatively reproduces several major morphological features of J0157, including the spiral gas structure and the diffuse bridge. These morphological similarities are consistent with a merger involving a substantial non-zero impact parameter, which can generate the observed large-scale spiral structure.
\\

\item Using GGM filtering and azimuthal surface-brightness analysis, we identify five quasi-periodic local maxima (A--E) along the front, interpreted as candidate KH rolls with a characteristic scale of $\sim$90 kpc --- consistent with estimates from the roll height, and comparable to KH scales reported in other clusters. 
The presence of these candidate KH rolls indicates that neither ICM viscosity nor a stabilizing magnetic layer fully suppresses instability growth at the interface. We place an upper limit on the effective viscosity of $f_\mu \lesssim 0.59$ times the Spitzer value. Alternatively, requiring that the KH growth timescale be shorter than the tidal timescale of the interaction, we derive an upper limit on the magnetic field strength at the front of $B \lesssim 5.9\ \mu$G. Both constraints are broadly consistent with values inferred for other well-studied CFs.
\end{itemize}

These results establish that J0157 provides a useful system for studying merger induced gas dynamics and for placing quantitative constraints on ICM microphysics, complementary to studies of classical sloshing CFs. Follow-up radio observations and higher-resolution, tailored magnetohydrodynamics simulations of this specific system would help to further test and refine the physical picture presented here.

\section*{Acknowledgements}
This research has made use of data and/or software provided by the High Energy Astrophysics Science Archive Research Center (HEASARC).
C.G. acknowledges support from the National Natural Science Foundation of China (Nos. 12422302, 12373007). L.J. acknowledges support from the NSFC (grant 12233005). G.L. acknowledges the research grants from the Ministry of Science and Technology of China (National Key Program for Science and Technology Research and Development, No. 2023YFA1608100), the research grants from the China Manned Space Project (CMS-CSST-2025-A08), and the National Natural Science Foundation of China (Nos. 12273036). Z.H. acknowledges support from the National Natural Science Foundation of China (12222304, 12192220, 12192221). H.X. acknowledges support from the National Natural Science Foundation of China No. 12233005. N.O. acknowledges JSPS KAKENHI grant number JP25K07368.

This research employs a list of \cha\ datasets, obtained by the \cha\ X-ray Observatory, contained in~\dataset[DOI: 10.25574/cdc.541]{https://doi.org/10.25574/cdc.541}.

\bibliography{ustc}{}
\bibliographystyle{aasjournalv7}

\end{document}